\documentstyle[aps,preprint]{revtex}
\begin{document}
\tighten
\newcommand{\mb}[1]{\mbox{\boldmath $#1$}}

\title{\bf \large
Sum Rule of the Hall Conductance
in Random Quantum Phase Transition}
\author{
Y. Hatsugai$^{1,2},$
K. Ishibashi$^1$ and
Y. Morita$^1$
}
\address{
 Department of Applied Physics,
 University of Tokyo,
 7-3-1 Hongo Bunkyo-ku, Tokyo 113, Japan$^1$ \\
 PRESTO, Japan Science and Technology Corporation$^{2}$
}
\maketitle

\begin{abstract}
The Hall conductance $\sigma_{xy}$
of two-dimensional {\it lattice} electrons
with
random potential
is investigated.
The change of $\sigma_{xy}$
due to randomness
is focused on.
It is a quantum phase transition
where the {\it sum rule} of $\sigma_{xy}$
plays an important role.
By the {\it string } (anyon) gauge,
numerical study becomes possible
in sufficiently weak magnetic field regime which 
is essential to discuss
the floating scenario in the continuum model.
Topological objects in the Bloch wavefunctions,
charged vortices,
are  obtained explicitly.
The anomalous plateau transitions
( $\Delta \sigma_{xy}= 2,3,\cdots >1$)
and the trajectory of delocalized states
are discussed.
\end{abstract}
\pacs{73.40.Hm Quantum Hall effect (integer and fractional)  
    73.50.-h Electronic transport phenomena in thin films and low-dimensional structures}
\vskip 1.0cm
\narrowtext

Effects of randomness
are crucial
in the quantum Hall effect (QHE)\cite{iqhe}.
According to the scaling theory
of the Anderson localization,
all states in two dimensions
are localized due to randomness\cite{aalr}.
There are, however,
some exceptions.
Symmetry effects,
which govern universality classes
of the Anderson localization,
allow the existence of delocalized states
in several two-dimensional systems.
For example, states at the center of each Landau band
become delocalized in the presence of strong magnetic field.
They are not extended in a usual manner,
but critical
which is associated
with multifractal character.
They play an essential role
in the quantization of the Hall conductance (QHE)
and their behavior determines the plateau transition in QHE.
The plateau transition
occurs as the strength of randomness or magnetic field is varied.
It is a typical quantum phase transition and
the Hall conductance characterizes each phase.

Non-zero Hall conductance means the existence
of delocalized states below the Fermi energy.
When the strength of randomness
is sufficiently strong,
the system is expected
to become the Anderson insulator.
Then it means disappearance of the
delocalized states below the Fermi energy.
If we assume that
the delocalized states do not disappear discontinuously,
they must float upward across the Fermi energy.
This is the floating scenario for
the delocalized states\cite{laughlin,khme}.
It is later extended
in the discussion of
the global phase diagram\cite{KLZ}.
The scenario also predicts
the selection rule
between different integer quantum Hall states.
The rule
prohibits transitions $\Delta \sigma_{xy} \neq \pm 1$.
On the other hand,
the breakdown of this selection rule is observed
in some experiments \cite{bd1a,bd1b,bd1c}
and numerical simulations\cite{sheng}.

In this paper,
based on topological arguments
of the Hall conductance\cite{tknn,yh}
and numerical study
of the lattice  model,
we try to clarify the points.
The plateau transition in QHE is a quantum phase transition
where the {\it sum rule}
restricts the transition type.
Therefore it is also interesting
as a problem of the quantum phase transition.
There are several studies on
the delocalized states in two-dimensional lattice electrons
with uniform magnetic field
and random potential\cite{sheng,ando1,mk1,yang,liu}.
Here we make clear
topological nature
of the Bloch wavefunctions
and the Hall conductance
in sufficiently weak magnetic field regime.
It has much to do with the continuum model and
may shed some light on the experiments\cite{bd1a,bd1b,bd1c}.
 The physical reason of the anomalous plateau transition is first
stated clearly in our paper.

The Hamiltonian
is defined
on a two-dimensional square lattice
as
\[
H=\sum_{\langle l,m\rangle} c_l^\dagger e^{i\theta_{lm}} c_m + h.c.
+\sum_{n}w_n  c_n^\dagger  c_n,\]
where
$c_n^\dagger(c_n)$
creates (annihilates)
an electron at a site $n$
and ${\langle l,m\rangle}$ denotes nearest-neighbor sites.
The magnetic flux per plaquette, $\phi$, is given by
$\sum_{\rm plaquette} \theta_{lm}=2\pi \phi$
where the summation runs over four links
around a plaquette.
The last term,  $w_{n}=Wf_{n}$,
is the strength of random potential
at a site $n$ and
$f_{n}$'s are uniform random numbers
between $[-1/2,1/2]$.
Although the system is infinite,
two-dimensional periodicity of $L_x\times L_y$
is imposed on $\theta_{lm}$ and $w_n$
(the infinite size limit corresponds to
$L_x, L_y{\rightarrow}\infty$).
When the randomness strength is weak and the temperature is sufficiently
low, the interactions between the electrons may play a dominant role.
However, we focus on the situation where it is not important and use this
non-interacting model.

When the Fermi energy lies
in the lowest $j$-th energy gap,
the Hall conductance $\sigma_{xy}$ is obtained
by summing
the Chern number $C_n$
below the Fermi energy
\[
\sigma_{xy}= \sum_{n=1}^j C_n
\]
\[
 C_n =\frac 1 {2\pi i}
\int d{\mb k}\; \hat z \cdot (\mb{ \nabla}_k \times{\mb A}_n), \; \; \;
 {\mb A}_n = \langle u_n (\mb{k}) | \mb{\nabla} _k | u_n(\mb{k}) \rangle,
\]
where
$| u_n(\mb{k}) \rangle$ is a Bloch wavefunction of the
$n$-th energy band with $L_xL_y$ components
and $u_n^\gamma(\mb{k})$ is  the $\gamma$-th component.
The integration  is over the Brillouin zone.
Arbitrarily
choosing the $\alpha$ and $\beta$-th components of the wavefunction and
focusing on
the winding number (vorticity  or charge)
at each zero point (vortex) of the $\alpha$-th component,
the expression is rewritten as
\[
C_n=\sum_\ell N_{n\ell},\;\;\; N_{n\ell}=\frac 1 {2\pi}
\oint_{\partial R_\ell}d\mb{k} \cdot \mb{\nabla} {\rm Im\  ln\ }
(\frac {u^\alpha_{n}(\mb{k})}  {u^\beta_{n}(\mb{k})}),
\]
where $N_{n\ell}$ is
the charge of a vortex at  $\mb{k}_\ell$
(a zero point of $u_n^{\alpha}(\mb{k})$ in the Brillouin zone),
$R_\ell$ is a region around $\mb{k}_\ell$
which does not
include other zero points of
the $\alpha$-th nor $\beta$-th components,
and $\partial R_\ell$ is the boundary.
The arbitrariness in the choice of $\alpha$ and $\beta$
corresponds to a freedom in gauge fixing.
In other words,
the gauge choice
does not affect observables such as the Hall conductance.
However, we need to fix the gauge to
obtain physical quantities.
Further, gauge dependent objects,
e.g. the configuration of vortices,
are helpful to understand the physics.
As discussed below,
the {\it sum rule} of the Hall conductance
can be clearly understood
by tracing the vortices.
Here we comment on
implication of the Chern number in the infinite size limit.
If all states in the $n$-th band are localized,
the corresponding Chern number vanishes, $C_n=0$.
Therefore non-zero $C_n$
means the existence of
delocalized states.
On the other hand, even if $C_n=0$,
we can not exclude
the existence of delocalized states
in a rigorous sense.
However, it is a situation in the Hall insulator and
we believe that
all the states in the band become localized as the Anderson insulator
of the orthogonal class (not unitary).

There are several previous studies on
the Hall conductance (Chern number)
in this model
\cite{sheng,yang}.
In this paper,
the plateau transition in
sufficiently weak magnetic field regime
is focused on
by the topological arguments and the numerical study.
It has much connection with
the continuum model.
In order to explore the regime,
the choice of the gauge is important.
There are several choices for a given geometry of the system.
Employing the novel {\it string } (anyon) gauge,
we can study topological nature
of the Bloch wavefunctions
in sufficiently weak magnetic field regime.
An example for ${\theta}_{ij}$'s in the string gauge
is shown for a $ 3\times 3$ square lattice
in Fig.\ref{fig:gauge}.
The extension to the other geometries is straightforward.
Choosing a plaquette $S$ as a starting one,
we  draw
outgoing arrows (strings)
from the plaquette $S$.
The $\theta_{ij}$  on a link $ij$
is given by $2\pi\phi n_{ij}$ where
$n_{ij}$ is the number of strings
which cut the link $ij$
(the orientation is taken into account).
Then it is clear that the magnetic flux is uniform
except at the plaquette $S$ .
At the plaquette $S$ , the condition of uniformity
gives $e^{-i2 \pi \phi (L_x L_y-1)}=e^{i2 \pi \phi }$.
It restricts the possible magnetic flux as
\[
\phi=\frac { n  }{L_xL_y}, \ \ n=1,2,\cdots, L_x L_y.
\]
In the case of the standard Laudau gauge in $x$-direction,
the smallest compatible magnetic flux $\phi$ with the periodicity
is   $\phi=1 /L_{x}$.
Then a system with a rectangular geometry is needed
for weak magnetic field regime.
On the other hand, in our string gauge,
it is $\phi=1 /L_xL_y$.
For a square $L\times L$ geometry, it allows us to make use of
$L$ times smaller magnetic flux in the string gauge
than the Landau gauge.
The string gauge enables us to study
sufficiently weak magnetic regime.

We performed numerical diagonalizations
of the above Hamiltonian and obtained
the spectrum and the Bloch wavefunctions.
Here the string gauge is employed.
Zero points (vortices)
of the Bloch wavefunctions
and the winding number (charge) of each vortex
are also calculated for all the energy bands.
In Figs.\ref{fig:chern-zero},
the configuration of vortices and their charges
are shown with
different
randomness strengths $W$'s.
As discussed above,
when the Fermi energy lies in an energy gap,
the Hall conductance is quantized to an integer.
The integers are shown in some energy gaps
in Figs.\ref{fig:chern-zero}.
They are given by the sum of
all the Chern numbers below each gap.
Note that,
although the energy gap has to close to change the Hall conductance
as discussed below,
the exact gap-closing points can not be seen
due to the lack of numerical accuracy.
However, tracing the vortices,
we can identify the gap closing points and
some of them are shown
by the triangles
in Figs.\ref{fig:chern-zero}.

Let us first summarize some features of the numerical results
in Figs.\ref{fig:chern-zero}.
As the strength of randomness $W$ is changed,
vortices in each energy band move continuously
and the motion of a vortex forms a {\it vortex line}.
However, with small change of $W$,
the Chern number of each energy band is stable.
This is
because the Chern number is
the {\it topological invariant}
of the energy band
and
the topology change
is necessary
to change it.
As seen in Figs.\ref{fig:chern-zero},
when the Chern number changes,
the two energy bands touch.
Then the two bands, in generic,
touch at one point in the Brillouin zone and
it forms a singularity.
Near the point,
the low-energy physics is described
generally by massless Dirac fermions\cite{NNN,osh1,NNN2}.
Associated with
the appearance of the gap closing point,
a vortex line passes through it and
the Chern number in each energy band changes by $\pm 1$.
Therefore
the Chern numbers of the two energy bands
do not change in total.
This is the {\it sum rule}
in our model\cite{NNN,osh1,NNN2,sum}.
It also leads to
the {\it selection rule } $\Delta \sigma_{xy}=\pm 1$.
As shown in Figs.\ref{fig:chern-zero},
the overlap of the energy bands can also happen.
Then the energy gap seems to be closed.
However there is still a energy gap in the Brillouin zone
(the situation is similar to semi-metals) and
the Chern number of each energy band is well-defined.
Therefore the vortex motion is still governed by
massless Dirac fermions.

As discussed above,
the basic observation of Figs.\ref{fig:chern-zero}
is that
the change of the Chern number
in each energy band
is described by massless Dirac fermions
and, in general, obeys the selection rule $\Delta \sigma_{xy}=\pm 1$.
In fact,
one can see the transition
$\sigma_{xy}=3{\rightarrow}2{\rightarrow}1{\rightarrow}0$
in Fig.\ref{fig:chern-zero} (a)
where the electron density is fixed.
However, the change of the {\it observed} $\sigma_{xy}$ 
can break the rule  (anomalous plateau transition).
The definition of the {\it observed} $\sigma_{xy}$ has subtle aspects. The
$\sigma_{xy}$  depends strongly on the randomness realization, the boundary
condition and the geometry in this transition region. Therefore naive
infinite size limit is not well-defined mathematically. One needs
to define the {\it observed} Hall conductance by its average over
different realizations. Then one can take the infinite size limit.
Physically, in a realistic situation, there are several possibilities for
the justification of the ensemble average. For example, (i) due to the
finite coherence length, the system effectively decouples into several
domains with different realizations of randomness or (ii) since the thermal
fluctuation exists, the $\sigma_{xy}$ is averaged near the Fermi energy and
the energy average may be replaced by the ensemble average \cite{meso}. 
In our model,
small energy gaps can appear
due to the existence of randomness.
It is clearly seen in Fig.\ref{fig:chern-zero} (b).
The Hall conductance is quantized
even when the Fermi energy lies in the small gap.
However these small gaps
strongly depend on randomness realization
and, after the ensemble average,
the corresponding Hall conductance
is generally not quantized. 
This is in contrast to the case
when the Fermi energy lies in the Landau gap
and the Hall conductance is quantized even after the ensemble average.
In fact,
after the ensemble average,
the plateau transitions
$\sigma_{xy}=3{\rightarrow}2{\rightarrow}1{\rightarrow}0$
in Fig.\ref{fig:chern-zero} (a) 
becomes $\sigma_{xy}=3\to0$ ($\Delta \sigma_{xy}=3$).
It demonstrates the transition
$\Delta \sigma_{xy}\neq \pm 1$ (See also Fig.\ref{fig:chern-av}).
Although
the plateau transitions generally obey
the selection rule ( $\Delta \sigma_{xy}=\pm 1$)
for a given realization of randomness,
the transition $\Delta \sigma_{xy}\neq \pm 1$
is observed due to the ensemble average.
This is the anomalous transition.

Finally we comment
on the trajectory of
delocalized states in the lowest Landau band.
As seen in Fig.\ref{fig:chern-zero} (b),
the lowest Landau band splits into some subbands
due to the existence of randomness.
In our case ($\phi=1/64$),
before the collapse of the lowest Landau gap,
the Hall conductance $\sigma_{xy}=1$
when the Fermi energy lies in the gap.
It implies that
there are delocalized states in the lowest Landau band.
In Fig.\ref{fig:chern-zero} (b),
when $W$ is sufficiently small,
only one of the subbands
in the lowest Landau band
carries non-zero Chern number (=+1).
Therefore
we can
assign
the position of the delocalized states
to the subband.
In this way,
the delocalized states can be traced\cite{com1}.
Through
the energy gap closing,
the Chern number changes
and the delocalized states move
from one subband to another.
Further,
when the gap closes,
the position of the delocalized states
can be identified with
the gap closing points i.e.
the massless Dirac fermions.
These gap closing points are shown
by the triangles.
It can be seen
in Fig.\ref{fig:chern-zero} (b) that,
when the randomness strength
is sufficiently small,
the delocalized states float up
relatively
within the lowest Landau band.
At the same time,
the lowest Landau band
broadens and goes downward in energy.
However,
before the delocalized states
float across the lowest Landau gap,
the gap collapses and
the states
disappear in pair with the other delocalized states
falling down from the higher energy region.
The pair annihilation point
depends strongly on the randomness realization
and the energy is not observed after the ensemble average
(it corresponds to the experimental situation).
It has the same origin
as the anomalous transition discussed above.
Therefore
we do not observe
any sign of floating across the Landau gap.

This work was supported in part by Grant-in-Aid
from the Ministry of Education, Science and Culture
of Japan and also Kawakami Memorial Foundation.
The computation  has been partly done
using the facilities of the Supercomputer Center,
ISSP, University of Tokyo.

\begin{figure}
\caption{
The definition
of the string (anyon) gauge
for a  $3 \times 3$ square system
with periodic boundary condition.
\label{fig:gauge}}
\end{figure}

\begin{figure}
\caption{
Zero points (vortices) of the Bloch wavefunctions
with their winding numbers (charges).
The shaded regions show energy bands.
Each circle denotes
the position of a vortex
as a function of randomness strength ($W$).
The full circle means the charge $+1$
and the open circle $-1$.
The Hall conductance is quantized to an integer,
when the Fermi energy lies at an energy gap.
The integers are shown at some gaps.
The gap closing points can be seen
within the numerical accuracy and some of them are shown
by triangles.
(a):
The  flux per plaquette
is $\phi=\frac 1 {64}$
and the system size is
$L_x \times L_y = 8 \times 8$.
(b):
The  flux per plaquette
is $\phi=\frac 1 {64}$
and the system size is
$L_x \times L_y = 24 \times 24$.
Only the region near the lowest Landau Level is shown.
\label{fig:chern-zero}}
\end{figure}

\begin{figure}
\caption{
The {\it observed } Hall conductance as a function of randomness strength
where the electron density is fixed. 
The flux per plaquette is $\phi=1/64$ and the system size is 
$L_x\times L_y=8\times 8$. 
The ensemble average 
is performed over 100 different realizations
of randomness. 
\label{fig:chern-av}}
\end{figure}


\begin{references}
\bibitem{meso}
On the other hand, when the temperature is sufficiently low that (i) the
coherence length grows and the $\sigma_{xy}$ for a given realization of
randomness is observed or (ii) the thermal fluctuation is below the level
spacing of the system, the $\sigma_{xy}$ generally depends strongly on the
randomness realization and the non-universal region should appear in the
plateau transition with the change of electron density. Although this
situation is not realistic in the macroscopic system, this non-universal
region may  appear in the mesoscopic system.
\bibitem{com1}
It is also possible that
the Chern numbers of neighboring subbands
change as $\{0,0\}\rightarrow\{+1,-1\}$.
It is due to the pair creation of vortices.
Then it seems
that there are several subbands
which include delocalized states.
In the case, the trajectory of the delocalized states
is complicated.
However, when the energy gap between the neighboring subbands
is much smaller than the randomness strength,
it is a finite size effect in general.
Therefore, after an ensemble average over
different realizations of randomness,
it may not be relevant.
\end{references}
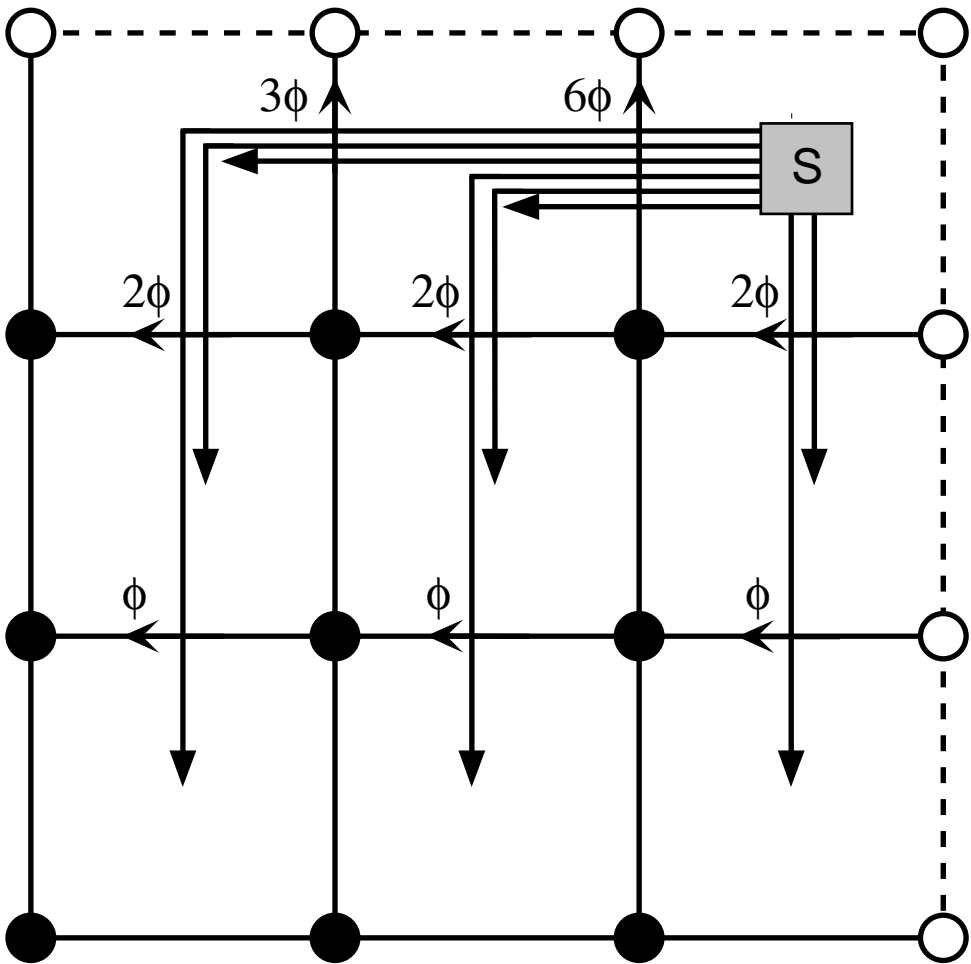
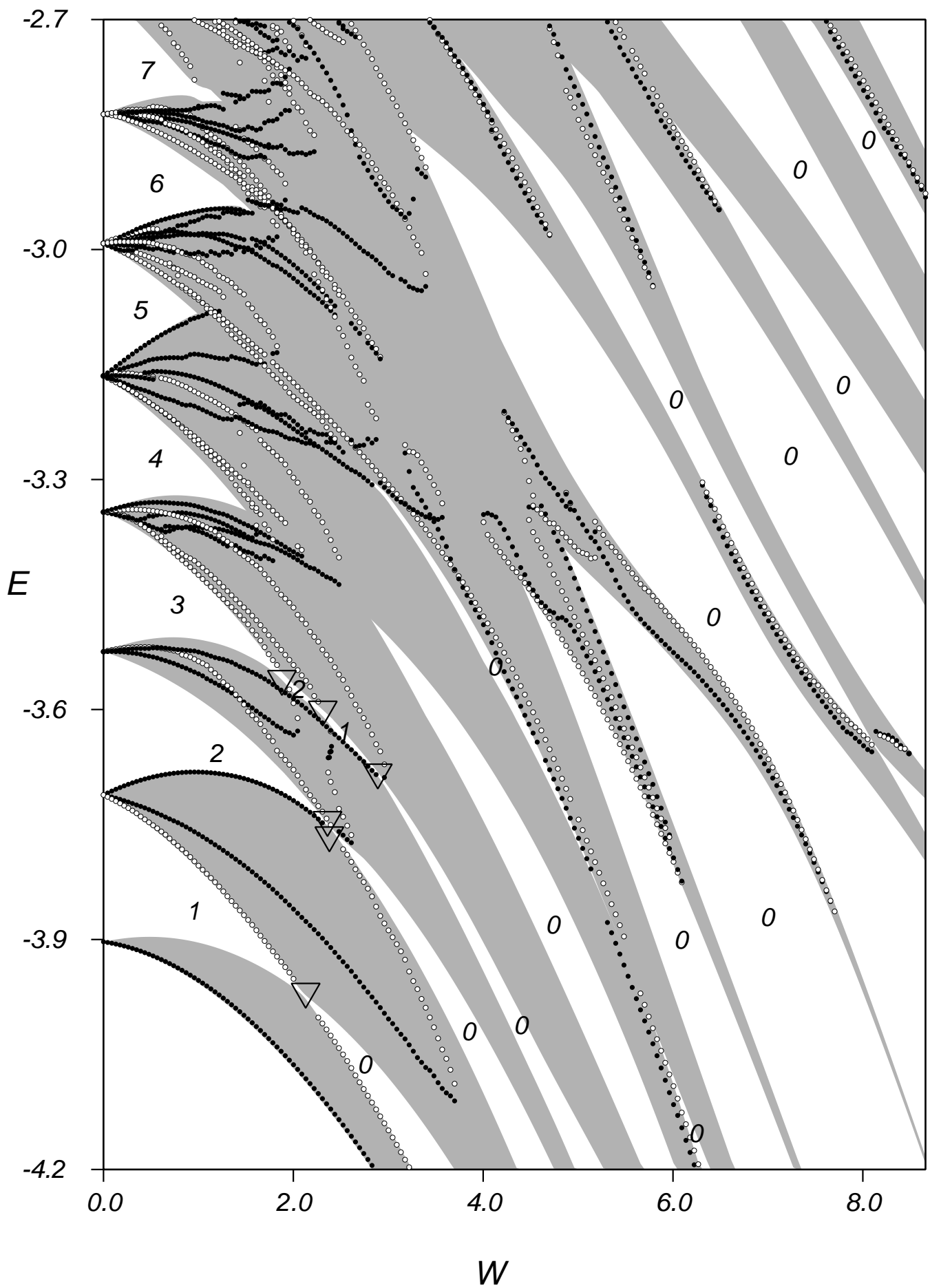
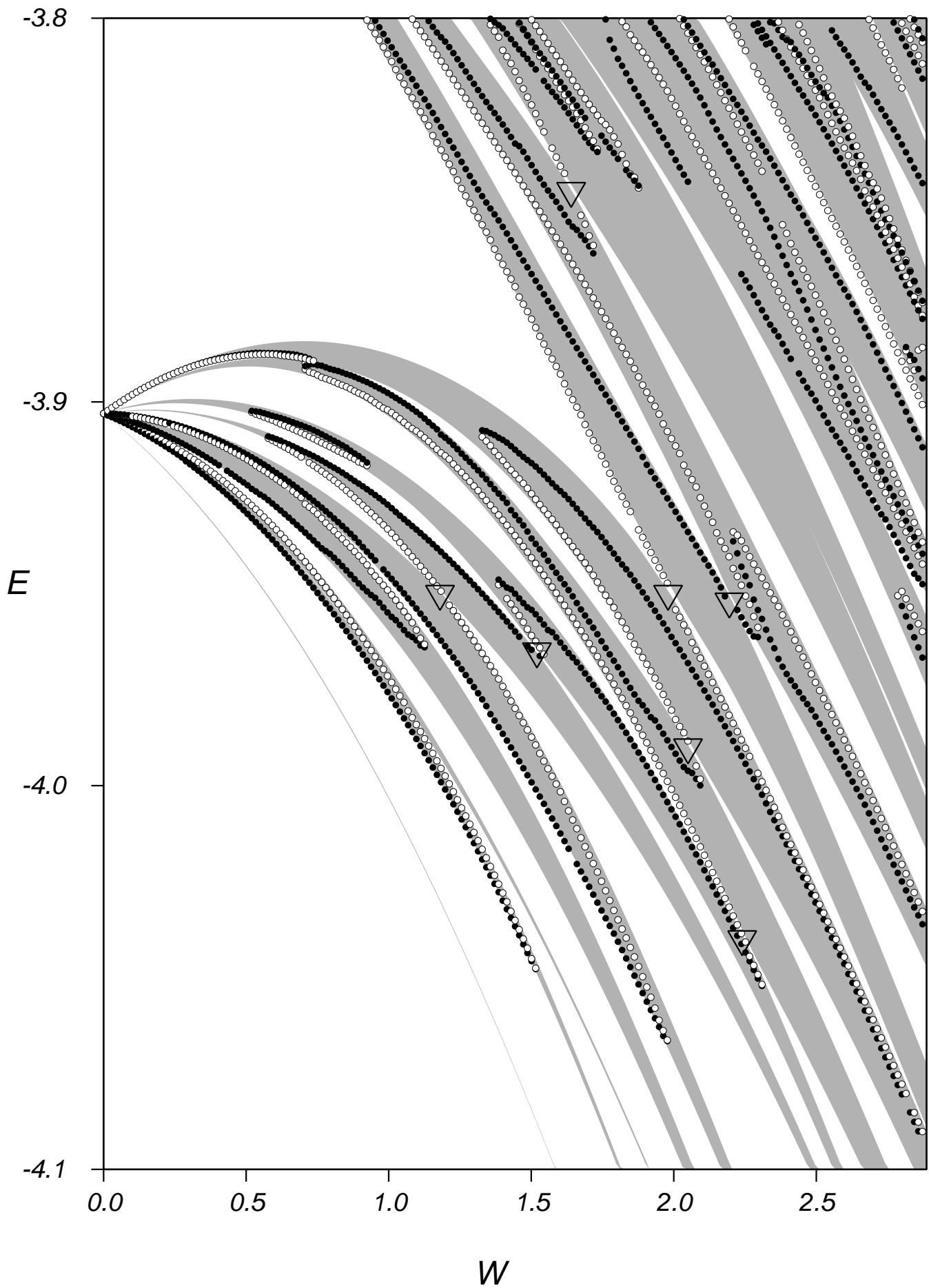
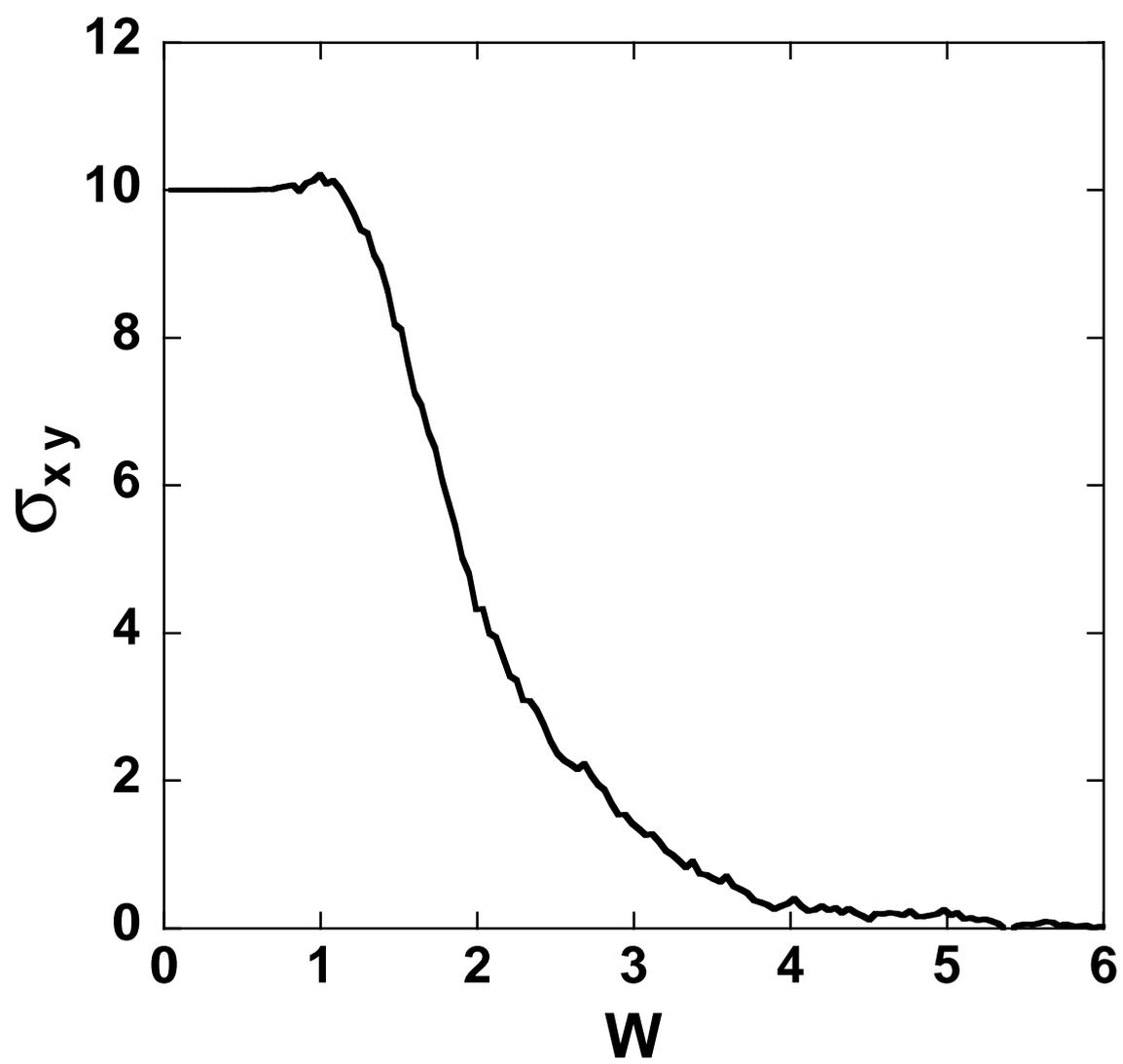
\end{document}